\documentclass[journal=jacsat,manuscript=article]{achemso}

\usepackage{graphicx}
\usepackage{dcolumn}
\usepackage{bm}
\usepackage{mathrsfs}
\usepackage[]{xcolor}
\usepackage{siunitx}
\usepackage{float}
\usepackage{amsmath,amssymb}
\usepackage{booktabs}
\usepackage{multirow}
\usepackage{pdfpages}
\usepackage[utf8]{inputenc}
\usepackage[english]{babel}

\usepackage{hyperref}
\usepackage[frozencache=true,cachedir=minted-output]{minted}
\usepackage{mdframed}

\SectionNumbersOn

\AtBeginEnvironment{minted}{\linespread{1.0}}

\definecolor{code}{rgb}{0.807843137254902, 0.8705882352941177, 0.9607843137254902}
\definecolor{output}{rgb}{0.8588235294117647, 0.9490196078431372, 0.8823529411764706}

\mdfdefinestyle{code_frame}{%
    backgroundcolor=code,
    linewidth=0.5pt,
}
\mdfdefinestyle{out_frame}{%
    backgroundcolor=output,
    linewidth=0.5pt,
}

\renewcommand{\mathbf}{\boldsymbol}

\DeclareUnicodeCharacter{2211}{$\Sigma$}
\DeclareUnicodeCharacter{3B4}{$\delta$}
\DeclareUnicodeCharacter{3C7}{$\chi$}
\DeclareUnicodeCharacter{393}{$\Gamma$}
\DeclareUnicodeCharacter{3B3}{$\gamma$}

\title{SpinAdaptedSecondQuantization.jl 1.0 - A Simple and Pedagogical Approach to Symbolic Quantum Chemistry}

\author{Marcus T.~Lexander}
\affiliation{%
 Department of Chemistry, Norwegian University of Science and Technology, 7491 Trondheim,
Norway
}
\author{Tor S.~Haugland}
\affiliation{%
 Department of Chemistry, Norwegian University of Science and Technology, 7491 Trondheim,
Norway
}
\author{Federico Rossi}
\affiliation{%
 Department of Chemistry, Norwegian University of Science and Technology, 7491 Trondheim,
Norway
}%
\author{Henrik Koch}
\email{henrik.koch@ntnu.no}
\affiliation{%
 Department of Chemistry, Norwegian University of Science and Technology, 7491 Trondheim,
Norway
}

\date{\today}

\begin{document}

\maketitle

\begin{abstract}
\noindent

The development of new electronic structure methods is a very time consuming and error prone process when done by hand. SpinAdaptedSecondQuantization.jl is an open-source Julia package we have developed for working with automated electronic structure theory development. The code focuses on being user-friendly and extensible, allowing for easy use of both user- and pre-defined fermionic and/or bosonic operators, tensors, and orbital spaces. This allows the code to be used to efficiently investigate and prototype new electronic structure methods for many different types of systems. This includes both exotic systems with wave functions consisting of different kinds of particles at once, as well as new parametrizations for traditional many electron systems. The code is spin-adapted, working directly with spin-adapted fermionic operators, and can easily be used to derive common electronic structure theory equations and expressions, such as the coupled cluster energy, ground and excited state equations, one- and two-electron density matrices, etc. Additionally, the code can translate expressions into code, accelerating the process of going from ideas to implemented methods.

\end{abstract}

\newpage

\section{Introduction}


%

When working with wave function based electronic structure methods, the second quantization framework is usually employed, as this allows for an efficient description of different operators and wave function parametrizations. Doing this by hand is, however, very time consuming and prone to human error. To address this, over the past many decades, there has been a lot of effort put into developing codes that assist in the derivation and implementation of various many-body methods, such as perturbation theory\cite{paldus_computer_1973,macleod_communication_2015} and coupled cluster theory\cite{cizek_explicit_1990,janssen_automated_1991,li_automation_1994,harris_computer_1999,nooijen_towards_2001,wladyslawski_analytical_2005,piecuch_automated_2006,datta_non-antisymmetric_2013,samanta_first-order_2018,quintero-monsebaiz_equation_2023,brandejs_generating_2025}.
Efforts have also been put towards the development of more generally useful algorithms and software packages for working with second quantization both using diagrammatic\cite{bochevarov_general_2004,datta_non-antisymmetric_2013} and direct symbolic\cite{hirata_tensor_2003,auer_automatic_2006,krupicka_toolchain_2017,rubin_pdaggerq_2021,evangelista_automatic_2022,liebenthal_automated_2025} approaches. 
These codes work by treating symbolic expressions of operators and tensors, allowing for the definition of common operators, such as the electronic Hamiltonian, as well as various wave function parametrizations. To obtain programmable expressions for equations, such as the coupled cluster equations, these expressions consisting of various operators need to be acted onto bra and ket states. Programmatically this is usually performed using an implementation of Wick's theorem \cite{wick_evaluation_1950} by forming a "normal ordering" of the electron annihilation and creation operators. As these operators reference spin orbitals directly, the expressions obtained from such codes need to be explicitly spin summed to obtain familiar spin adapted expressions such as the coupled cluster equations derived in the "Molecular Electronic Structure Theory" book\cite{helgaker_molecular_2013}. In this paper we present the 1.0 release of the open source \mintinline{julia}{SpinAdaptedSecondQuantization.jl} package, in which we work directly with the high level spin-adapted operators which makes the derivation stay much closer to what one would do by hand. This means that common intermediates, such as commutators and projections can be more easily understood and worked with, either manually or programmatically. The package is written in pure Julia\cite{bezanson_julia_2017} which makes it easy to work with in an interactive manner using a combination of Julia scripts and the interactive Julia Read-Eval-Print Loop (REPL), allowing for rapid exploration and prototyping, assisting in the derivation of new methods. The package is also highly extensible, allowing users to easily define new operator types, tensor types, and orbital/index spaces in your own input scripts that stand on equal footing to the internal types provided in the package. The package is easily available through the Julia package manager, and has already been used in multiple published works\cite{lexander_analytical_2024,castagnola_strong_2024,castagnola_realistic_2025,folkestad2025excitation,rossi_generalized_2025} as well as being used for the current implementation of the QED-CCSD and CCSDT methods in the eT program\cite{folkestad_et_2020}.

\section{Second Quantization}


In quantum chemistry one usually works in the framework of Second Quantization to describe the many-body wave functions of electronic systems. The most basic building blocks in this framework are the fermionic annihilation and creation operators, denoted by the symbols $a$ and $a^\dagger$ respectively. These operators are associated with a certain spin-orbital, as they represent removing or adding an electron to it. If, for example, we start with the vacuum state $|\text{vac}\rangle$ and act the creation operator $a^\dagger_{1\alpha}$ on it, we end up with the one-electron wave function $a^\dagger_{1\alpha} |\text{vac}\rangle$ with a single spin-up electron in orbital number 1. Further if we act another creation operator, $a^\dagger_{1\beta}$ we get the two-electron wave function $a^\dagger_{1\beta} a^\dagger_{1\alpha} |\text{vac}\rangle$ consisting of the single Slater determinant with two electrons of opposite spin, both in orbital number 1.
Using this, we can describe a closed shell Hartree-Fock determinant as applying creation operators for all the occupied spin-orbitals on the vacuum state
\begin{equation}
    |\text{HF}\rangle = \left(\prod_{i}{a^\dagger_{i \beta} a^\dagger_{i \alpha}}\right) |\text{vac}\rangle
\end{equation}
where we use the indices $i, j, \dots$ to denote occupied orbitals. A defining algebraic property of the creation operators is that they anti-commute, that is, their anti-commutator is zero
\begin{equation}
    [a^\dagger_P, a^\dagger_Q]_+ = a^\dagger_P a^\dagger_Q + a^\dagger_Q a^\dagger_P = 0.
\end{equation}
A direct consequence of this is that the antisymmetry in the Slater-determinant is baked into the operators as
\begin{equation}
    a^\dagger_P a^\dagger_Q |\Phi\rangle = -a^\dagger_Q a^\dagger_P |\Phi\rangle.
\end{equation}
Here we have used capital indices $P, Q, \dots$ to denote general spin orbitals. As mentioned, the adjoint of the creation operator is the annihilation operator, which removes an electron from its spin orbital and as it is the adjoint of the creation operator, the same commutation properties apply. Another defining property of these operators is their shared anticommutation relation
\begin{equation}
    [a_P, a^\dagger_Q]_+ = \delta_{PQ}
\end{equation}
with $\delta_{PQ}$ being the Kronecker delta
\begin{equation}
    \delta_{PQ} = \left\{\begin{array}{lc}
        1 & P = Q \\
        0 & P \neq Q
    \end{array}\right.
\end{equation}
Usually in quantum chemistry, we are working with an N-electron wave function of a certain spin symmetry which leads us to introducing the singlet excitation operator
\begin{equation}
    E_{pq} = a^\dagger_{p \alpha} a_{q \alpha} + a^\dagger_{p \beta} a_{q \beta},
\end{equation}
where we have used indices $p, q, \dots$ to denote spatial orbitals.
The action of these operators is to move an electron from orbital $q$ to orbital $p$ and retaining the spin symmetry.
Now we can also concisely express the molecular Hamiltonian in its second quantization representation as
\begin{equation}
    H = \sum_{pq}{h_{pq} E_{pq}}
    +
    \frac12 \sum_{pqrs}{g_{pqrs} e_{pqrs}}
    +
    h_{nuc}
\end{equation}
where we have defined the two-electron singlet excitation operator
\begin{equation}
    e_{pqrs} = \sum_{\sigma \tau}{a^\dagger_{p\sigma} a^\dagger_{r\tau} a_{s\tau} a_{q\sigma}}
    =
    E_{pq} E_{rs} - \delta_{qr} E_{ps}
\end{equation}
and the one- and two-electron integrals are given by
\begin{equation}
    h_{pq} = \langle\phi_p| \hat h |\phi_q\rangle
\end{equation}
\begin{equation}
    g_{pqrs} = (pq|rs)
\end{equation}
as well as the nuclear repulsion given by $h_{nuc}$.

\noindent
Using commutation relations for the annihilation/creation operators, one can derive the commutation relation for the excitation operators
\begin{equation}
    [E_{pq}, E_{rs}] = \delta_{qr} E_{ps} - \delta_{ps} E_{rq}.
\end{equation}
Since this commutator is expressed only in terms of excitation operators themselves and the terms all have only one operator, instead of the two you would get from naively multiplying out the commutator, it is often advantageous to work directly with these operators instead of the annihilation/creation operators.

\section{The Challenge}

The second quantization formalism described above gives us a systematic way of working with electronic structure methods, however, even for the quite simple standard models, such as CCSD, the number of intermediates one has to deal with when deriving ground and excited state equations can be quite large. As a result, the derivation and subsequent implementation of new electronic structure methods is a long and tedious process when done by hand, usually lasting for weeks, to obtain even just a proof of concept implementation. The effort required increases rapidly with the increased complexity of the system Hamiltonian, and wave function parametrizations. This makes many potentially useful methods completely out of reach for humans to derive and implement, as putting many months or years of effort towards such methods is impractical.

Fortunately, the very tedious and time consuming process of deriving second quantization expressions is very systematic and only boils down to a few basic steps. A lot of the time is spent performing commutators between second quantization operators, and automating only this step would save a lot of time in itself. Consider taking the commutator between two expressions consisting of $M$ and $N$ terms each,
\begin{align}
    A &= A_1 + A_2 + ... + A_M,
    \\
    B &= B_1 + B_2 + ... + B_N,
\end{align}
with each of the terms, $A_I$ and $B_J$, respectively consisting of a coefficient $C_I$ and $D_J$ and a string of operators $a_{I,i}$ and $b_{J,j}$,
\begin{align}
    A_I &= C_I a_{I,1} a_{I,2} ... a_{I,m_{I}},
    \\
    B_J &= D_J b_{J,1} b_{J,2} ... b_{J,n_{J}}.
\end{align}
Now consider the commutator $[A, B]$. The first step is to expand the bilinearity of the commutator,
\begin{equation}
    [A, B] = [A_1, B_1] + [A_2, B_1] + ... + [A_1, B_2] + ... + [A_M, B_N]
    =
    \sum_{IJ}{[A_I, B_J]},
\end{equation}
resulting in $M \cdot N$ commutators between single terms. Now consider one of these commutators,
\begin{align}
    [A_I, B_J] = C_I [a_{I,1} a_{I,2} ..., B_J]
    &=
    C_I \left([a_{I,1}, B_J] a_{I,2} ... a_{I,m_{I}} + a_{I,1} [a_{I,2}, B_J] a_{I,3} ... a_{I,m_{I}} + ...\right),
    \\
    [a_{I,i}, B_J] = D_J [a_{I,i}, b_{J,1} ...]
    &=
    D_J \left([a_{I,i}, b_{J,1}] b_{J,2} ... b_{J,n_{J}} + b_{J,1} [a_{I,i}, b_{J,2}] b_{J,3} ... b_{J,n_{J}} + ...\right),
\end{align}
which results in $m_I \cdot n_J$ commutators between single operators. We can thus express the total number of basic commutators required to evaluate as,
\begin{equation}
    N_{\text{comm}} = \sum_{IJ}{m_I n_J},
\end{equation}
which grows very quickly. Further, commutators often show up inside other commutators in nested commutators, such as the ones arising from the Baker-Campbell-Hausdorff (BCH) expansion,
\begin{equation}
    e^{-B} A e^{B} = A + [A, B] + \frac12 [[A, B], B] + \frac{1}{3!} [[[A, B], B], B] + ...,
\end{equation}
quickly leading to an amount of terms that is impractical to deal with by hand. In the following sections, we describe our software package which automates the process of working with general second quantization expressions, including the commutator procedure described above.

\section{Structure of the Code}

\noindent
The main building block of the code is the Term structure. A Term contains an array of Kronecker deltas, an array of tensors and an array of operators. This lets us represent terms like $\delta_{pr} g_{pqrs} E_{pq} E_{rs}$. In addition to this, the Term contains an array of which its indices are summed over and a map of index constraints. This allows us to restrict certain indices to belong to a certain subspace of orbitals, most commonly the occupied or virtual spaces. Finally the Term contains a scalar factor, so we can have terms like
\begin{equation}
    \frac12 \sum_{aibj}{t_{aibj} E_{ai} E_{bj}}.
\end{equation}
The scalar type can be any numerical type, however it is best to avoid floating point types as round-off errors can lead to terms not canceling properly.

\noindent
The Expression structure is the building block which one mostly interacts directly with.
This is mostly just an array of Terms which the Expression keeps sorted and will fuse together Terms that are equal up to the scalar in front.

\subsection{Simplification}

\noindent
The Terms contain the main simplification functionality. The most basic kind of simplification is recognizing summation indices which show up in a Kronecker delta. This allows simplification like the following
\begin{equation}
    \sum_{pqrs}{g_{pqrs} \delta_{qr} E_{ps}} = \sum_{pqs}{g_{pqqs} E_{ps}}
\end{equation}

\noindent
Another type of simplification that the Term can do is the renaming of indices. There are two main cases where indices can be renamed, one being summation indices, which can be renamed to any free index name, and the other case being indices that show up in Kronecker deltas. For example a term like $\delta_{pq} E_{pq}$ could be rewritten as $\delta_{pq} E_{pp}$. When simplifying, a Term will try to rename indices in order to minimize them. For Kronecker deltas, this means renaming any index in the Term that shows up in the delta to the lowest index in the delta, while for summation indices, the lowest free index names will be used. After lowering the summation indices there is one more step we can take to further simplify the Term, which is to reorder the summation indices. Take for example the term $\sum_{pq}{h_{qp} E_{qp}}$. If we reorder the indices $p, q \rightarrow q, p$ we get the term $\sum_{pq}{h_{pq} E_{pq}}$ which looks better as it has a lower lexicographical ordering. Finding the permutation of the summation indices that gives the lowest lexicographical ordering is often necessary in order for equal Terms to cancel (or combine), but it is not always trivial to find the best ordering. In the example above, it is possible to find the best ordering as we can take the order the indices first show up in the Term and sort this to find the permutation. However, if we use tensors with some symmetry like $h_{pq} = h_{qp}$ and revisit the Term from before, since the tensor is symmetric, it will already have reordered its indices to minimize its ordering and we would have the term $\sum_{pq}{h_{pq} E_{qp}}$. We see here that the order the indices first show up will always be $p,q$ so to find the best ordering of the indices for the Term we are forced to check both permutations. This procedure scales as the factorial of the number of summation indices, which already for 8 indices is over 40 thousand, but still this is usually not the bottleneck.

\subsection{The "Reductive" Commutator}

\noindent
Evaluating commutators between Terms consisting of many operators is essential for deriving many expressions showing up in electronic structure methods, especially when performing projections and expectation values. This is often performed by repeated application of the following identities
\begin{equation}
    [A B, C] = A [B, C] + [A, C] B,
\end{equation}
\begin{equation}
    [A, B C] = [A, B] C + B [A, C].
\end{equation}
For operator types that reduce rank when commuting, these relations work well. For different types of operators, however, like fermionic annihilation and creation operators, it might be needed to perform anti-commutators rather than commutators. Further, when commuting arbitrary strings of operators, the choice of whether the outer commutation should be a commutator or an anti-commutator depends on the specific commutation of pairs of operators. We define a "reductive" commutator
\begin{equation}
    [A, B]_\Gamma = A B + \Gamma \cdot B A, \qquad \Gamma \in \{1, -1\},
\end{equation}
where we have introduced the sign constant $\Gamma$ to denote whether it is a commutator or anti-commutator. We can then derive similar commutation identities to the ones above
\begin{equation}
    [A B, C]_\Gamma = A [B, C]_{\Gamma_1} - \Gamma_1 \cdot [A, C]_{\Gamma_2} B,
    \qquad
    \Gamma = -\Gamma_1 \Gamma_2,
\end{equation}
\begin{equation}
    [A, B C]_\Gamma = [A, B]_{\Gamma_1} C - \Gamma_1 \cdot B [A, C]_{\Gamma_2},
    \qquad
    \Gamma = -\Gamma_1 \Gamma_2,
\end{equation}
where the outer commutator sign $\Gamma$ is determined from the signs for the sub-commutators $\Gamma_1$ and $\Gamma_2$ which are in turn determined to produce Terms with the least operator rank. These identities are then repeatedly employed until commutators between single operators are reached, keeping track of the signs of each sub-commutator accumulating the final sign of the outer commutator. The commutation relations of each pair of operator types are then implemented in the code, returning both the commutator expression and the sign $\Gamma$ to signal whether that particular pair of operator types commute or anti-commute.

\subsection{The "act\_on\_ket" Function}

\noindent
When evaluating projections and expectation values, most codes resort to implementing Wick's theorem, and express all operators in terms of ferminonic creation and annihilation operators. However, when working by hand, one usually works without expanding operators in terms of the basic ladder operators, rather working with the commutation relations of the operators to move certain operators to project on the HF bra and ket to simplify the expression as much as possible. Our code follows a similar procedure which can be described as follows. For each operator type we implement an "act\_on\_ket" function which works as a base case for the procedure, the job of which is to simplify the projection as much as possible by introducing constraints on indices and reducing the number of operators. A good example is the implementation of the "act\_on\_ket" function for the $E_{pq}$ operator,
\begin{equation}
    E_{pq} |\text{HF}\rangle = \left\{\begin{array}{ll}
        2 \delta_{pq} |\text{HF}\rangle & p, q\ \text{both occupied} \\
        E_{pq} |\text{HF}\rangle & p\ \text{virtual}, q\ \text{occupied}
    \end{array}\right.,
    \label{eq:Epq_example}
\end{equation}
where we see that we have two non-zero cases. In both cases, the index $q$ gets restricted to refer to an occupied orbital, while the two cases are distinguished by whether the index $p$ is occupied or virtual. The main point of the function is that it transforms the generic operator $E_{pq}$ into two terms, where one represents a non-excited determinant, and the other represents a singly excited determinant. Now, in order to do the same for a string of many operators, we have the following procedure. We write the string of operators acting on a HF-ket as
\begin{equation}
    A_1 A_2 ... A_{n-1} A_n |\text{HF}\rangle.
\end{equation}
For each of the basic operators $A_i$ the base case implementation of the act\_on\_ket function exists, such that we can write
\begin{equation}
    A_i |\text{HF}\rangle = \tilde A_i |\text{HF}\rangle,
\end{equation}
where the tilde $(\tilde A_i)$ represents the operator having been simplified according to its act\_on\_ket implementation, like for $E_{pq}$ in Eq. \ref{eq:Epq_example}. For the full string we can now write
\begin{equation}
\begin{split}
    A_1 A_2 \dots A_{n-1} A_n |\text{HF}\rangle
    &=
    A_1 A_2 \dots A_{n-1} \tilde A_n |\text{HF}\rangle
    \\
    =
    -\Gamma \cdot \tilde A_n \left(A_1 A_2 \dots A_{n-1} |\text{HF}\rangle\right)
    &+
    [A_1 A_2 \dots A_{n-1}, \tilde A_n]_\Gamma |\text{HF}\rangle,
\end{split}
\label{eq:act_on_ket}
\end{equation}
where we have reduced the full problem of $n$ operators acting on the ket into two smaller sub-problems, each with a maximum of $n-1$ operators acting on the ket. The first of these arises from having moved the last operator $\tilde A_n$ to the back and acting the remaining operators on the ket where we have picked up a potential sign change ($-\Gamma$) from reordering the operators. The second term is then the commutator that ensures we have not changed the total expression, which is chosen to be the reductive commutator described above, giving us both an expression with a lower operator rank, as well as the sign for the first term. Then Eq. \ref{eq:act_on_ket} is applied recursively on the sub-problems, eventually terminating when all branches are out of operators. After performing the full projection, we will have obtained an expression clearly separated into different classes of terms where, in particular, the terms with no operators left can be read out as the expectation value of the initial term. Further, we can read out the terms containing a single excitation operator as the surviving terms of a projection on a singly excited bra, and so on.

\section{Installation and Usage of the Package}

\noindent
In this section we illustrate the capabilities of the package with some examples of how the code can be used to derive spin-adapted equations for some familiar methods. For further details and examples the reader is referred to the package documentation available at \url{https://marcustl12.github.io/SpinAdaptedSecondQuantization.jl/}. The latest version of the package can easily be installed by running the following commands in a Julia terminal and typing a single \mintinline{julia}{]} to open the package mode then writing \mintinline{julia}{add SpinAdaptedSecondQuantization} and pressing enter.
\begin{mdframed}[style=code_frame]
\begin{minted}{julia}
julia> ]add SpinAdaptedSecondQuantization
\end{minted}
\end{mdframed}
To install the version 1.0 released with this paper, add \mintinline{julia}{@1.0} to the end when installing
\begin{mdframed}[style=code_frame]
\begin{minted}{julia}
julia> ]add SpinAdaptedSecondQuantization@1.0
\end{minted}
\end{mdframed}
When the package is successfully installed, it can be used by adding the line
\begin{mdframed}[style=code_frame]
\begin{minted}{julia}
using SpinAdaptedSecondQuantization
\end{minted}
\end{mdframed}
to your input script.

\subsection{Hamiltonian and HF energy expression}

We can define the one-electron Hamiltonian operator

\begin{equation}
    h = \sum_{pq}{h_{pq} E_{pq}}
\end{equation}
with the following line of code
\begin{mdframed}[style=code_frame]
\begin{minted}{julia}
h = ∑(real_tensor("h", 1, 2) * E(1, 2) * electron(1, 2), [1, 2])
\end{minted}
\end{mdframed}
Breaking down the different parts here, we first have the \mintinline{julia}{real_tensor("h", 1, 2)} function. This produces the most basic type of tensor, with a name "h", and indices 1 and 2. Next is the \mintinline{julia}{E(1, 2)} function which produces a singlet excitation operator with indices 1 and 2. The function \mintinline{julia}{electron(1, 2)} produces a term constraining indices 1 and 2 to the "GeneralOrbital" index space which is meant for general electronic indices. The final part is the ∑ function which produces a summation of the terms over indices 1 and 2. The unicode symbol ∑ can be written in a Julia terminal/REPL or editor by writing "\textbackslash sum" followed by a tab. Alternatively one can use the alias "summation" for the same function.

\noindent
Similarly we can define the two electron operator
\begin{equation}
    g = \frac12 \sum_{pqrs}{g_{pqrs} e_{pqrs}}
\end{equation}
with the code
\begin{mdframed}[style=code_frame]
\begin{minted}{julia}
g = 1//2 * ∑(psym_tensor("g", 1:4...) * e(1:4...) * electron(1:4...), 1:4)
\end{minted}
\end{mdframed}
The \mintinline{julia}{1//2} here is a rational fraction of integers. We discourage the use of floating point numbers such as 0.5 as rounding errors might lead to certain terms not canceling properly. The two electron excitation operator \mintinline{julia}{e(p, q, r, s)} is just an alias for the expression in terms of one electron operators
\begin{mdframed}[style=code_frame]
\begin{minted}{julia}
e(p, q, r, s) = E(p, q) * E(r, s) - δ(r, q) * E(p, s)
\end{minted}
\end{mdframed}
instead of being its own operator type. We can now get the HF energy expression by writing
\begin{mdframed}[style=code_frame]
\begin{minted}{julia}
E_HF = simplify(hf_expectation value(h + g))
println(E_HF)
\end{minted}
\end{mdframed}
which would output the following
\begin{mdframed}[style=out_frame]
\begin{minted}{julia}
2 ∑_i(h_ii)
+ 2 ∑_ij(g_iijj)
- ∑_ij(g_ijji)
\end{minted}
\end{mdframed}
Here we see a pattern that shows up a lot, the $2 g_{pqrs} - g_{psrq}$ pattern.
For the two electron integrals we define the tensor
\begin{equation}
    L_{pqrs} = 2 g_{pqrs} - g_{psrq}.
\end{equation}
To automatically recognize and simplify when this pattern occurs in an expression we can use the following
\begin{mdframed}[style=code_frame]
\begin{minted}{julia}
E_HF = look_for_tensor_replacements(E_HF,
    make_exchange_transformer("g", "L"))
println(E_HF)
\end{minted}
\end{mdframed}
which now would output
\begin{mdframed}[style=out_frame]
\begin{minted}{julia}
2 ∑_i(h_ii)
+ ∑_ij(L_iijj)
\end{minted}
\end{mdframed}

\subsection{Tensors and Symmetry}

In the above definitions of the Hamiltonian operator and HF energy expressions we have used two different tensor types, the \mintinline{julia}{real_tensor} and the \mintinline{julia}{psym_tensor}. The first of these represents the most basic type of tensor, an n-index array of real numbers with no assumed symmetry and should be the default choice in cases where one wants to derive general expressions without imposing any symmetry on integrals and coefficients. The \mintinline{julia}{psym_tensor} represents even number indexed tensors that have particle-exchange symmetry, such as the two-electron integrals, $g_{pqrs}$, and coupled cluster amplitudes, $t_{ij...}^{ab...}$. This symmetry is implemented by having the tensor sort the pairs of neighboring indices. This, for example, gives 4-index tensors such as $g_{pqrs}$ the two-fold symmetry,
\begin{equation}
    g_{pqrs} = g_{rspq},
\end{equation}
6-index tensors with 6-fold symmetry,
\begin{equation}
    t_{aibjck} =
    t_{aickbj} =
    t_{bjaick} =
    t_{bjckai} =
    t_{ckaibj} =
    t_{ckbjai},
\end{equation}
and in general, $2n$-index tensors has $n!$-fold symmetry. In addition to these two tensor types, we have the \mintinline{julia}{rsym_tensor} type, which gives $2n$-indexed tensors the full $n!\cdot2^n$-fold symmetry which $n$-electron integral matrices of real orbitals would have. For example the two-electron integrals would have the full 8-fold symmetry,
\begin{equation}
    g_{pqrs} =
    g_{pqsr} =
    g_{qprs} =
    g_{qpsr} =
    g_{rspq} =
    g_{srpq} =
    g_{rsqp} =
    g_{srqp},
\end{equation}
when represented with this type of tensor. If other types of symmetric tensors are required they can be implemented as user defined tensor types in the input script, with the given symmetry. Detailed instructions are given in the package documentation.

\subsection{CCSD equations}

When working with coupled cluster expressions we usually deal with the Fock matrix $F_{pq}$ rather than the one-electron matrix $h_{pq}$. For this reason we usually express the Hamiltonian in terms of the Fock matrix using the following code
\begin{mdframed}[style=code_frame]
\begin{minted}{julia}
h = ∑((real_tensor("F", 1, 2) +
    ∑((-2 * psym_tensor("g", 1, 2, 3, 3) +
            psym_tensor("g", 1, 3, 3, 2), [3])
)) * E(1, 2) * electron(1, 2), [1, 2])
\end{minted}
\end{mdframed}
Next we need to define our cluster operator which for this example will be the $T_2$ operator only. This is because the $T_1$ operator can be included in the Hamiltonian by adjusting the one- and two-electron integrals and are therefore rarely included when deriving equations.\cite{helgaker_molecular_2013} In doing this we implicitly assume that the Hamiltonian is written in terms of the $T_1$ transformed integrals, which have less symmetry than the standard integrals, and it important that we define the Hamiltonian using the \mintinline{julia}{real_tensor} and the \mintinline{julia}{psym_tensor} tensor types which do not assume more symmetries than the $T_1$ transformed integrals have. We define the $T_2$ operator with the following code
\begin{mdframed}[style=code_frame]
\begin{minted}{julia}
T2 = 1//2 * ∑(psym_tensor("t", 1, 2, 3, 4) * E(1, 2) * E(3, 4) *
    occupied(2, 4) * virtual(1, 3), 1:4)
\end{minted}
\end{mdframed}
Now we can derive the CCSD equations. We start by deriving the expression for the similarity transformed Hamiltonian $\bar H = e^{-T} H e^T$ by using the BCH expansion
\begin{equation}
    \bar H = e^{-T} H e^T
    =
    H + [H, T] + \frac12 [[H, T], T] + \dots
\end{equation}
For the electronic Hamiltonian and cluster operator this truncates after only five terms. We can perform this using the following code
\begin{mdframed}[style=code_frame]
\begin{minted}{julia}
Hbar = bch(H, T2, 4)
\end{minted}
\end{mdframed}
where the number 4 indicates that we only want to include terms up to 4th order from the BCH expansion, which is the last non-zero term.
The next step is to project onto the HF ket to obtain $\bar H |\text{HF}\rangle$
\begin{mdframed}[style=code_frame]
\begin{minted}{julia}
Hbar_ket = simplify(act_on_ket(Hbar, 2))
\end{minted}
\end{mdframed}
The number 2 is to discard any terms that have more than two operators remaining which are not needed for any CCSD expression as we will be projecting on no more than doubly excited left states. This can significantly speed up the projection.
We can now project on various bra-states to obtain different useful quantities.

\subsubsection{Energy}
The simplest quantity is the energy which we can obtain by projecting on the HF bra. Subtracting off the HF energy, we can get an expression for the correlation energy as
\begin{mdframed}[style=code_frame]
\begin{minted}{julia}
E_CCSD = act_on_bra(Hbar_ket)
E_HF = hf_expectation_value(H)
E_corr = simplify_heavy(E_CCSD - E_HF)
E_corr = look_for_tensor_replacements(
    E_corr, make_exchange_transformer("t", "u"))
println(E_corr)
\end{minted}
\end{mdframed}
which would output
\begin{mdframed}[style=out_frame]
\begin{minted}{julia}
∑_iajb(g_iajb u_aibj)
\end{minted}
\end{mdframed}
Here we have used the \mintinline{julia}{look_for_tensor_replacements} function to simplify the expression using the relation
\begin{equation}
    u_{aibj} = 2 t_{aibj} - t_{ajbi}.
    \label{eq:t2u}
\end{equation}

\subsubsection{Singles}
Next is to derive the expression for the singles part of the CCSD equations
\begin{equation}
    \Omega_{ai} = \tilde{\left\langle_i^a\right|} \bar H |\text{HF}\rangle = 0,
    \label{eq:omegaai}
\end{equation}
Here the state $\tilde{\left\langle_i^a\right|}$ is the biorthonormal state to the ket state $\left|_i^a\right\rangle = E_{ai} |\text{HF}\rangle$ defined such that
\begin{equation}
    \left\langle\tilde{_i^a}| _j^b\right\rangle
    =
    \delta_{ij} \delta_{ab},
\end{equation}
which for single excitations can be explicitly achieved by
\begin{equation}
    \tilde{\left\langle_i^a\right|} = \frac12 \langle\text{HF}| E_{ia}.
\end{equation}
To compute the projection in Eq. \ref{eq:omegaai} we could either explicitly multiply the $E_{ia}$ on the left and project, or we can directly insert Kronecker deltas from the definition of the biorthonormality. This can be done by running the following
\begin{mdframed}[style=code_frame]
\begin{minted}{julia}
template_ket = E(1, 2) * occupied(2) * virtual(1)
omega_ai = project_biorthogonal(Hbar_ket, template_ket)
omega_ai = look_for_tensor_replacements(
    omega_ai, make_exchange_transformer("t", "u"))
println(omega_ai)
\end{minted}
\end{mdframed}
which would output
\begin{mdframed}[style=out_frame]
\begin{minted}{julia}
F_ai
+ ∑_jb(F_jb u_aibj)
+ ∑_bjc(g_abjc u_bicj)
- ∑_jkb(g_jikb u_ajbk)
\end{minted}
\end{mdframed}
The \mintinline{julia}{project_biorthogonal(...)} function takes in a template ket which is used instead of an explicit expression for the biorthogonal bra.

\subsubsection{Doubles}
\label{sec:doubles}
Similarly we can obtain an expression for the doubles part of the CCSD equations
\begin{equation}
    \Omega_{aibj} = \tilde{\left\langle_{ij}^{ab}\right|} \bar H |\text{HF}\rangle.
\end{equation}
Here the biorthonormal bra, $\tilde{\left\langle_{ij}^{ab}\right|}$, is defined in terms of a biorthogonal bra, $\bar{\left\langle_{ij}^{ab}\right|}$,\cite{helgaker_molecular_2013}
\begin{equation}
    \tilde{\left\langle_{ij}^{ab}\right|}
    =
    \frac{1}{1 + \delta_{ab}\delta_{ij}}
    \bar{\left\langle_{ij}^{ab}\right|},
\end{equation}
where the biorthogonal bra is defined such that the following relation holds
\begin{equation}
    \bar{\left\langle_{ij}^{ab}\right|}\left. _{kl}^{cd}\right\rangle
    =
    P_{ij}^{ab} \delta_{ac} \delta_{bd} \delta_{ik} \delta_{jl},
    \label{eq:biortho_2}
\end{equation}
where the permutation operator $P_{ij}^{ab}$ generates the sum of all permutations of the pairs $(a, i)$ and $(b, j)$. The biorthogonal bra is possible to define explicitly as
\begin{equation}
    \bar{\left\langle_{ij}^{ab}\right|}
    =
    \langle\text{HF}| \left(\frac13 E_{ia} E_{jb} + \frac16 E_{ja} E_{ib}\right),
\end{equation}
however, for triple excitations and higher it is not possible to express the biorthogonal basis in terms of singlet excitation operators. We use the definition of the biorthogonality in Eq. \ref{eq:biortho_2} to project using the following code
\begin{mdframed}[style=code_frame]
\begin{minted}{julia}
omega_aibj = project_biorthogonal(Hbar_ket,
    E(1, 2) * E(3, 4) * occupied(2, 4) * virtual(1, 3))
omega_aibj = simplify_heavy(symmetrize(omega_aibj,
    make_permutation_mappings([(1, 2), (3, 4)])))
omega_aibj = look_for_tensor_replacements(omega_aibj,
    make_exchange_transformer("t", "u"))
\end{minted}
\end{mdframed}
The call here to the \mintinline{julia}{symmetrize(...)} function performs the expansion of the permutation operator from the definition of the biorthogonality condition in Eq. \ref{eq:biortho_2} which is required in order to properly simplify the expression. The expression we have produced now will include a lot of redundant terms that arise from the symmetry of the expression, for example including both of the terms
\begin{equation}
    \sum_c{F_{ac} t_{bjci}} +
    \sum_c{F_{bc} t_{aicj}},
    \label{eq:redundant}
\end{equation}
however, only one of these is required to compute if the resulting tensor is symmetrized numerically by adding its transpose to itself. To remove the redundant terms we can run the following
\begin{mdframed}[style=code_frame]
\begin{minted}{julia}
omega_aibj_redundant,
omega_aibj_self_symmetric,
omega_aibj_non_symmetric =
    desymmetrize(omega_aibj,
    make_permutation_mappings([(1, 2), (3, 4)]))
\end{minted}
\end{mdframed}
This function returns three expressions, the first of which are the terms where the "mirrored" counterparts have been removed, and that needs to be symmetrized to recover the full symmetric output. This expression would, for example, contain only one of the two terms in \ref{eq:redundant}. The second expression are the "self-symmetric" terms, which are themselves symmetric under the given permutations and can be evaluated without any extra steps. The third expression contains any leftover terms which were not found to be either symmetric nor have a symmetric counterpart somewhere in the full expression. When computing something that has a known symmetry, like the doubles part of the CCSD equations, this should ideally be zero.
Outputting the separate parts of this, we get the following for the redundant
\begin{mdframed}[style=code_frame]
\begin{minted}{julia}
println(omega_aibj_redundant)
\end{minted}
\end{mdframed}
\begin{mdframed}[style=out_frame]
\begin{minted}{julia}
∑_c(F_ac t_bjci)
- ∑_k(F_ki t_akbj)
- ∑_ck(g_acki t_bjck)
- ∑_ck(g_ackj t_bkci)
+ ∑_kc(g_aikc u_bjck)
- ∑_kcld(g_kcld t_aibk u_cjdl)
- ∑_kcld(g_kcld t_aicj u_bkdl)
\end{minted}
\end{mdframed}
the self-symmetric
\begin{mdframed}[style=code_frame]
\begin{minted}{julia}
println(omega_aibj_self_symmetric)
\end{minted}
\end{mdframed}
\begin{mdframed}[style=out_frame]
\begin{minted}{julia}
g_aibj
+ ∑_cd(g_acbd t_cidj)
+ ∑_kl(g_kilj t_akbl)
+ ∑_kcld(g_kcld t_akbl t_cidj)
+ ∑_kcld(g_kcld t_akdj t_blci)
+ ∑_kcld(g_kcld u_aick u_bjdl)
\end{minted}
\end{mdframed}
and finally the non symmetric
\begin{mdframed}[style=code_frame]
\begin{minted}{julia}
println(omega_aibj_non_symmetric)
\end{minted}
\end{mdframed}
\begin{mdframed}[style=out_frame]
\begin{minted}{julia}
∑_kcld(g_kcld t_aicl t_bkdj)
- ∑_kcld(g_kcld t_bjcl u_aidk)
\end{minted}
\end{mdframed}
Here we see that the non-symmetric terms are in fact not zero, even though the total output should be symmetric by construction. This is a rare case where the previous simplification using the relation in Eq. \ref{eq:t2u} has made the symmetry harder to recognize, though if one re-expands the latter of the terms in terms of $t$ instead of $u$ the symmetry can be easily verified.
In order to evaluate the doubles omega we have to evaluate and symmetrize the redundant terms, then add the symmetric terms (including the seemingly non-symmetric terms) which we can write like
\begin{equation}
    \tilde \Omega_{aibj} = P_{ij}^{ab} \Omega_{aibj}^r + \Omega_{aibj}^s,
\end{equation}
where the superscripts $r$ and $s$ denote the "redundant" and "symmetric" terms respectively, with the redundant needing to be symmetrized to include the missing redundant terms. The tilde on $\tilde \Omega$ denotes that the expression was derived using the biorthogonal rather than the biorthonormal bra-states and thus needs to be scaled by a factor of $\frac12$ on the diagonal
\begin{equation}
    \Omega_{aibj} = \frac{1}{1 + \delta_{ij} \delta_{ab}} \tilde \Omega_{aibj}.
\end{equation}

\newpage

\subsection{Generation of Numerical Code}

After having generated expressions for the quantities like the CCSD omega equations, we want to be able to
generate runnable code we could use as part of an implementation of the method we are deriving. The expressions generated by the code are already very close in syntax to the "einsum" syntax of packages like numpy. We provide a function that translates an expression into a runnable Julia function that uses the einsum notation of the \mintinline{julia}{TensorOperations.jl}\cite{TensorOperations.jl} package. As an example we can use the \mintinline{julia}{omega_aibj_redundant} expression from above. Running the code
\begin{mdframed}[style=code_frame]
\begin{minted}{julia}
function_string = print_julia_function(
    "compute_omega_r", omega_aibj_redundant; 
    explicit_tensor_blocks = ["t", "u"])
println(function_string)
\end{minted}
\end{mdframed}
would output the following
\begin{mdframed}[style=out_frame]
\begin{minted}{julia}
function omega_doubles_r(no, nv, F, t_vovo, g, u_vovo)
    nao = no + nv
    o = 1:no
    v = no+1:nao
    
    X = zeros(nv, no, nv, no)
    g_vvoo = @view g[v,v,o,o]
    g_voov = @view g[v,o,o,v]
    g_ovov = @view g[o,v,o,v]
    F_vv = @view F[v,v]
    F_oo = @view F[o,o]
    @tensoropt (a=>10χ,b=>10χ,c=>10χ,d=>10χ,i=>χ,j=>χ,k=>χ,l=>χ) begin
        X[a,i,b,j] += F_vv[a,c]*t_vovo[b,j,c,i]
        X[a,i,b,j] += -F_oo[k,i]*t_vovo[a,k,b,j]
        X[a,i,b,j] += -g_vvoo[a,c,k,i]*t_vovo[b,j,c,k]
        X[a,i,b,j] += -g_vvoo[a,c,k,j]*t_vovo[b,k,c,i]
        X[a,i,b,j] += g_voov[a,i,k,c]*u_vovo[b,j,c,k]
        X[a,i,b,j] += -g_ovov[k,c,l,d]*t_vovo[a,i,b,k]*u_vovo[c,j,d,l]
        X[a,i,b,j] += -g_ovov[k,c,l,d]*t_vovo[a,i,c,j]*u_vovo[b,k,d,l]
    end
    X
end
\end{minted}
\end{mdframed}
Here we have specified that sub-blocks of the tensors $t$ and $u$ should be explicit input parameters to the function as there is only one of them (the $t_{aibj}$ block) while the Fock matrix and electron repulsion integrals are taken in as full tensors with indices running over all molecular orbitals and having the function extract sub-blocks.

\subsection{Fermions and Bosons}

Since the code works directly with any operator type, it is easy to work with expressions using both fermionic and bosonic operators together, such as the bilinear part of the Pauli-Fierz Hamiltonian used for QED-CCSD\cite{haugland_coupled_2020}
\begin{equation}
    H_{\text{bilinear}}
    =
    \sum_{pq}{d_{pq} E_{pq} (b^\dagger + b)},
\end{equation}
where $b^\dagger$ and $b$ are respectively creation and annihilation operators for a photon of a given cavity mode. 
We can express this easily with the following code
\begin{mdframed}[style=code_frame]
\begin{minted}{julia}
H_bilinear = ∑(
    real_tensor("d", 1, 2) * E(1, 2) * (bosondag() + boson()) *
    electron(1, 2), 1:2)
\end{minted}
\end{mdframed}
Similarly we can also express the additional contributions to the cluster operator used in QED-CCSD, namely
\begin{align}
\Gamma &= \gamma b^\dagger,
\\
S_1 &= \sum_{ai}{s_{ai} E_{ai} b^\dagger},
\\
S_2 &= \frac12 \sum_{aibj}{s_{aibj} E_{ai} E_{bj} b^\dagger},
\end{align}
which can be expressed in code as
\begin{mdframed}[style=code_frame]
\begin{minted}{julia}
Γ = real_tensor("γ") * bosondag()
S1 = ∑(real_tensor("s", 1, 2) * E(1, 2) * bosondag() *
    occupied(2) * virtual(1), 1:2)
S2 = 1//2 * ∑(psym_tensor("s", 1:4...) *
    E(1, 2) * E(3, 4) * bosondag() *
   occupied(2, 4) * virtual(1, 3), 1:4)
\end{minted}
\end{mdframed}
We can now for example derive the bilinear terms for the omega singles equations of QED-CCSD
\begin{equation}
    \Omega_{ai}^\text{0} = \langle_i^a, 0| e^{-T} H_\text{bilinear} e^T |\text{HF}\rangle,
\end{equation}
using the code
\begin{mdframed}[style=code_frame]
\begin{minted}{julia}
Hbar = bch(H_bilinear, [T2, Γ, S1, S2], 4)
Hbar_ket = act_on_ket(Hbar, 3)
omega_ai = project_biorthogonal(
    Hbar_ket, E(1, 2) * occupied(2) * virtual(1)
omega_ai = look_for_tensor_replacements(
    omega_ai, make_exchange_transformer("t", "u"))
println(omega_ai)
\end{minted}
\end{mdframed}
which would output
\begin{mdframed}[style=out_frame]
\begin{minted}{julia}
d_ai γ
+ ∑_b(d_ab s_bi)
- ∑_j(d_ji s_aj)
+ 2 ∑_j(d_jj s_ai)
+ 2 ∑_jb(d_jb s_aibj)
- ∑_jb(d_jb s_ajbi)
+ ∑_jb(d_jb u_aibj γ)
\end{minted}
\end{mdframed}

Similarly one could derive expressions for the remaining parts of the ground state QED-CCSD equations as well as Jacobian transformation and density matrices. The implementation of QED-CCSD in the eT program\cite{folkestad_et_2020} including ground state molecular gradients\cite{lexander_analytical_2024} is almost fully based on autogenerated expressions obtained using SpinAdaptedSecondQuantization.

\newpage

\section{Timings}

Here we present a few timings to illustrate how far up the coupled cluster hierarchy we can go. In table \ref{tab:omega_timings} is a detailed breakdown of the timings of the different parts that go into deriving the coupled cluster ground state equations for truncation orders up to 6. The benchmarking script that was used to produce these timings is available in the \href{https://marcustl12.github.io/SpinAdaptedSecondQuantization.jl/v1.0/cc_bench/}{Coupled Cluster Benchmark} section in the documentation and is run with 112 threads on an Intel(R) Xeon(R) Platinum 8480+ dual socket system.

\begin{table}[!htbp]
    \centering
    \begin{tabular}{l|rrrr|r|c}
        \toprule
        Derivation Step & BCH       & Simplify  & Project   & Finalize  & Total     & New Terms \\
        \midrule
        CCS             & 14 µs     & 30.3 ms   & 26.5 ms   & 125 ms    & 0.2 s     & 1         \\
        CCSD            & 253 ms    & 31 ms     & 8.6 ms    & 454 ms    & 0.7 s     & 21        \\
        CCSDT           & 1.4 s     & 0.5 s     & 0.6 s     & 0.8 s     & 3.4 s     & 40        \\
        CCSDTQ          & 9.0 s     & 5.2 s     & 6.6 s     & 1.5 s     & 22.4 s    & 66        \\
        CCSDTQP         & 45.9 s    & 30.5 s    & 42.1 s    & 7.5 s     & 126 s     & 93        \\
        CCSDTQPH        & 198 s     & 139 s     & 274 s     & 105 s     & 717 s     & 133       \\
        CCSDTQPH7       & 694 s     & 511 s     & 1792 s    & 2375 s    & 5373 s    & 173       \\
        CCSDTQPH78      & 1663 s    & 1194 s    & 8353 s    & 55855 s   & 67065 s   & 228       \\
        \bottomrule
    \end{tabular}
    \caption{Timings for deriving expressions for ground state CC equations at different levels of truncation. All expressions are derived using cluster operator without $T_1$. The "Finalize" step includes simplifying, projecting on the biorthogonal basis and the subsequent simplifications from the \hyperref[sec:doubles]{CCSD example} above.}
    \label{tab:omega_timings}
\end{table}

\newpage

\section{Conclusion and Further Work}

We have presented the 1.0 release of the Julia package \mintinline{julia}{SpinAdaptedSecondQuantization.jl} for working with symbolic spin-adapted quantum chemistry methods. The package has proved very useful and has thus seen use for the development of various methods such as QED-CCSD and CCSDT. The code works directly with spin adapted operators, which together with the interactive nature of the package makes the process of deriving and developing new methods stay close to how one would derive similar methods by hand. This makes intermediates like transformed and projected operators easier to understand and reason about. The package is highly extensible, allowing for easy additions of new operator types and index spaces, making the package a useful tool when exploring the development of methods for new and exotic systems.

Currently the code is mainly focused on the development of coupled cluster methods, where in the future we would like to enhance its capabilities in working with different types of methodologies such as unitary transformations, response theory and multi-configurational reference wave functions. With the code providing many general tools for working with second quantization, we are confident that it is suitable for such further developments.
The code being written in the Julia programming language provides a powerful user and developer experience with quick and easy prototyping of new derivation strategies using code in the users own input files.

The code provides some simple functionality of translating expressions into numerical code, which is very powerful for the rapid prototyping of new methods. In the future we would like to expand these capabilities by both generalizing the code generation to allow for expressions containing Kronecker deltas as well as generating more optimal code that is able to isolate common intermediates, factorize expressions and exploit the symmetries of tensors. This would be done through a combination of interfacing to existing software that achieve similar functionality for tensor contractions, as well as implementing new tools where needed.

\newpage

\section{Supporting Information and Code Availability}

The newest version of the code and documentation is available \href{https://github.com/MarcusTL12/SpinAdaptedSecondQuantization.jl/}{on github}\cite{SpinAdaptedSecondQuantization.jl}, with the 1.0 version released with this paper is available at \\ \url{https://github.com/MarcusTL12/SpinAdaptedSecondQuantization.jl/releases/tag/v1.0.0} and can also be downloaded at the Zenodo repository\cite{zenodo}.

\noindent
For supporting information the reader is referred to version 1.0 of the package documentation available \url{https://marcustl12.github.io/SpinAdaptedSecondQuantization.jl/v1.0/}.

\section{Acknowledgements}

We acknowledge and thank Alexander Paul for helpful discussions and contributions early in the project. We also thank the following users of the code, M. Castagnola, S. D. Folkestad, K. L. Kruken, D. Lipovec, Y. el Moutaoukal, J. Pedersen, R. R. Riso, B. S. Sannes, W. Smedsrud, L. Stoll and J. H. M. Trabski for valuable testing and feedback.
We acknowledge funding from the European Research Council (ERC) under the European Union’s Horizon 2020 Research and Innovation Programme (grant agreement No. 101020016).

\newpage

\bibliography{references}

\section{ToC}

\includegraphics[width=\textwidth]{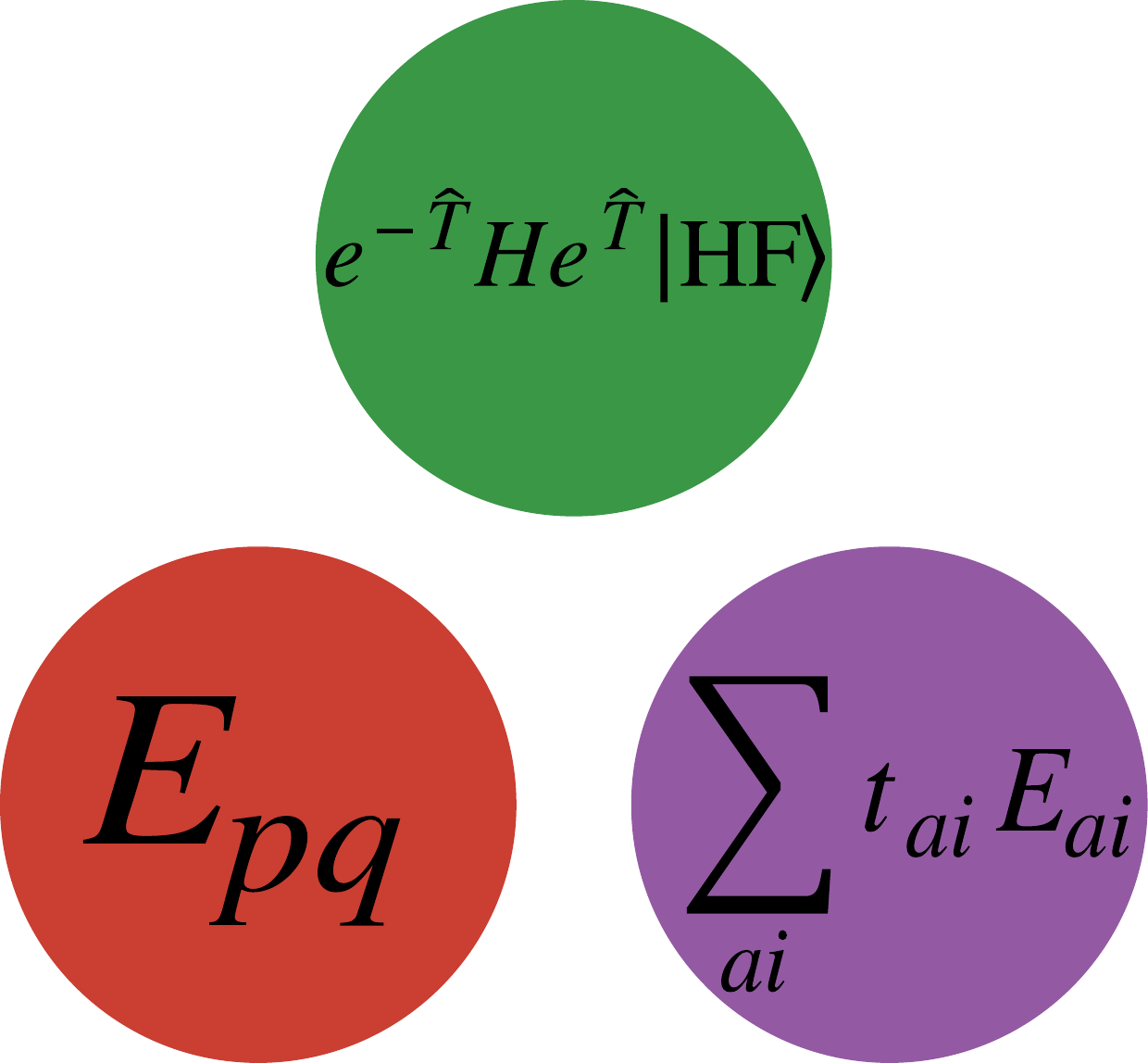}

\end{document}